\def\la{\mathrel{\mathpalette\fun <}}

\def\fun#1#2{\lower3.6pt\vbox{\baselineskip0pt\lineskip.9pt
  \ialign{$\mathsurround=0pt#1\hfil##\hfil$\crcr#2\crcr\sim\crcr}}}
\relax
\documentstyle[12pt]{article}
\advance\textheight by \topskip
\def\Box{{\,\lower0.9pt\vbox{\hrule \hbox{\vrule height 0.2 cm \hskip
0.2 cm \vrule height 0.2 cm}\hrule}\,}}
\begin{document}
\newcommand{\mn}{{\em Mon. Not. R. astr. Soc}}
\newcommand{\apj}{{\em Astrophys. J.}}
\newcommand{\aj}{{\em Astron. J.}}
\renewcommand{\aa}{{\em Astr. Astrophys.}}
\newcommand{\ass}{{\em Astrophys. Space Sci.}}
\newcommand{\nat}{{\em Nature}}
\newcommand{\et}{{\em et al.}\,\,}
\def\la{\langle}
\def\ra{\rangle}
\def\ti{\tilde}
\def\lam{\lambda}
\def\Ps{{\cal P}}
\def\Qs{{\cal Q}}
\def\rhos{\varrho}
\def\vs{\vartheta}
\def\vvs{\pmb{$\vartheta$}}
\def\Vs{{\cal V}}
\begin{titlepage}
\begin{flushright}
November 1993
\end{flushright}
\vskip 2.5truecm

\centerline {\bf COSMOLOGICAL   DISTRIBUTION FUNCTIONS $^{*}$}
\vskip 1.2cm
\centerline{Lev Kofman$^{1,2}$}

\vskip .5cm
\centerline{$^{1}$ Institute for Astronomy, University of Hawaii, HI 96822}
\centerline{$^{2}$ CIAR Cosmology program and CITA, U. of  Toronto, Canada}
\vskip 1.5truecm
\centerline{\bf Abstract}
\medskip
The evolution of probability distribution functions (PDFs) of continuous
  density, velocity
and velocity derivatives  ( deformation tensor) fields in the theory of
  cosmological gravitational
 instability
are considered.
We show that in the Newtonian theory the dynamical equations cannot
be reduced to  the closed set of Lagrangian equations.
 Since continuous fields from galaxy surveys need sufficiently
large  smoothing which exceeds the scale of nonlinearity, one can use
 the  Zel'dovich approximation to describe the mildly non-linear matter
 evolution, which allows  the closed set of Lagrangian equations.
The closed kinetic equation for the joint PDF
of cosmological continuous fields is derived in this approximation.
 The analytical theory of the cosmological PDFs
 with  arbitrary (including  Gaussian) initial
 statistics  is developed, based on the solution on the  kinetic
equaiton.\\
For Gaussian initial fluctuations,  the PDFs
are parametrized by only linear {\it rms}
fluctuations $\sigma$  on given filtering scale.
Density PDF  $P(\rho, t)$ and PDF
 $M(\lam_1, \lam_2, \lam_3; t)$ of eigenvalues of the deformation tensor
field
  in the Eulerian space  evolve
 very rapidly  in non-linear regime.  On the contrary, velocity PDF
 $Q(\vec v, t)$ remains invariant under non-linear evolution.
For small $\sigma$ the Edgworth series is suggested to reconstruct
 $P(\rho, t)$.
Density PDF $P(\rho, t)$ is close to the {\it log-normal} PDF for the CDM model
at moderate  $\sigma$, but differs from that in general case.

\vskip 2truecm

$^{*}$ Extended version of the Talk presented
 at the 9-th IAP Meeting ``Cosmic velocity field'',
Paris, June 12-17, 1993.

\end{titlepage}
\vfill
\eject

\section{Introduction}

One of the standard metods to study the statistical property of the non-linear
Large Scale Structure (LSS) of the universe is the two-point correlation
function $<y(\vec r_1) y(\vec r_2)>=\xi^{(2)}_y(\vec r_1, \vec r_2)$
 and further, n--point functions
$\xi_y^{(n)}(\vec r_1, ...,\vec r_n)$ of the point-like statistical ensemble of
galaxies, $y$ may be the density, velocity components or other fields.
 The full statistics of the ensemble is described by
the n--point PDFs $ P(y_1, ..., y_n)$ which are connected with
$\xi_y^{(n)}$  \cite{MY}.
It is insufficient to know a few lower order cerrelation functions of
galaxy distribution to define the full
 statistics of non-linear matter distribution
in the universe.
 In this contribution   we will  consider
 the simplest one--point
PDFs of the continuous non-linear cosmic fields.
On the statistics of cosmological point processes see \cite{BE}.\\
The theoretical  development  of the subject begun with
the calculations of the lowest moments $<\delta^{(n)}>$ of the density fields
 in the perturbation theory \cite{Peeb},\cite{Fry},
in the Zel'dovich approximation \cite{Gor},
for  different cosmologies \cite{Bouch}, for filtered field
\cite{JBC}, in numerical simulations for  non-linear regime  \cite{CF}.\\
Recently even more attention has been  drawn to the PDFs which can probe
the statistics of primordial fluctuations.   The standard CDM model with
the scale free Gaussian initial fluctuations is hardly compatible with
the set of observations.
 One  way  (fortunately not the only) is to switch the model
 to start with non-Gaussian primordial inhomogeneities. In the cosmic
 inflation scenario in principle, there is some room to introduce non-Gaussian
fluctuations  \cite{Kof3}, although  it needs fine tuning in
parameters,  as a rule.
Non-Gaussian perturbations also arise in other scenarios, where
progenitors of inhomogeneities are
 topological defects, such as textures
\cite{T}, cosmic strings \cite{Br},  or
explosions \cite{O}.\\
 The phenomenological attempts to design the density PDF at non-linear
 stage were made from different  arguments.
In    \cite{S} the  formula for $P(\rho)$
from ``thermodynamical''  treatment of gravitational system
was suggested, which evokes some sceptizism.
The log-normal mapping in order to mimick the non-linear density evolution
 was argued in  \cite{CJ}. Unfortunately the log-normal model
does not work \cite{CMS}, although  the
features of the log-normal distribution have been recognized in
 the density PDF   \cite{Ham}, \cite{CJ}, \cite{Kof2}.
 An elegant analytical method  has been developed  \cite{Ber}
 to summarize the
series of moments  for the small density fluctuations  $\delta$ in the
 approximation when
$<\delta^{(n)}> \approx S_n \sigma^{2(n-1)}$.
 In \cite{Kof},
 \cite{Kof2} we   calculated  $P(\rho)$
and $F(\vec v)$ analytically in the  ``truncated'' Zel'dovich approximation
(i.e. with initial smoothing),
 and found a good agreement
with the results of N-body simulations (with final smoothing).
Recently two different methods from  papers \cite{Ber}
and \cite{Kof2} were compared  resulting
in a   good  mutual agreement  \cite{BK}.\\
In \cite{PS} further approximation was made
 within  the Zel'dovich theory to calculate
 $P(\rho)$ after the final smoothing. Unfortunately the assumptions
 made are valid
for small fraction of the Lagrangian volume and  result does not fit
the numerical calculations.
The count-in-cell statistics which is tighly related to the $P(\rho)$
was calculated numerically in \cite{BH}.
In \cite{CW}  $P(\rho)$ from  a range of
numerical simulations with different  statistics was plotted.\\
 From observational side, the  skewness measured from the IRAS QDOT \cite{Sau}
 is in the range $s/\sigma =1 - 2$, as expected from
the Gaussian initial statistics. However, the IRAS galaxies are
 underrepresented
in clusters, which correspond to high density tail of $P(\rho)$,
contributing to the skewness. This has little effect on $P(\rho)$
itself because clusters occupy a small fraction of volume.
The density and velociy PDF as deduced from 1.9 Jy IRAS survey and
 POTENT analysis  is consisient with Gaussian initial statistics
\cite{Kof2}, the  count-in-cell analysis of 1.2 Jy IRAS data
also consistent with this hypothesis \cite{BDS}, although in all cases
the larger
sample volume is needed for more quantitative conclusion.\\

\section{ Basic Non-Local Equations}

At the epoch of the LSS formation, most of the mass is in the
 form of dark matter
of relic origin (for instance,  the Cold Dark Matter), with no pressure;
nonlinear LSS originated by gravitational instability of small
initial fluctuations.
Let $\vec x,  \vec v =a{d\vec x \over d t},  \rho  (t, \vec x)$ and
$\phi (t,\vec x)$ be, respectively, the comoving coordinates, peculiar velocity
and density of dark matter (neglecting baryons), and peculiar Newtonian
gravitational potential. In the Newtonian theory, the motion of  dark matter
 before the
particle orbits cross obeys a nonlinear system of equations
\begin{equation}
{\partial \rho \over \partial t} +3H\rho +
 {1\over a}\nabla_x(\rho \vec v)=0,
\label{eq:cont}
\end{equation}
\begin{equation}
{\partial \vec v \over \partial t} +
 {1 \over a}(\vec v \nabla_x )\vec v +H\vec v = -{1 \over a}
\nabla_x \phi, \label{eq:euler}
\end{equation}
\begin{equation}
{\nabla_x}^2\phi = 4\pi G a^2 (\rho - \bar \rho),    \label{eq:poiss}
\end{equation}
where $a(t)$ is a scalar factor of an expanding Universe, $H=\dot a/ a$ and
$\bar \rho$ is a mean density.
These equations are valid in the single stream regime, and admit the evident
generalization in the regions of multiple streams.\\
For simplicity we  assume that
 Inflation produces a flat universe $\Omega= 1$, and $\Lambda=0$.
The growing mode of adiabatic perturbations  $D(t)$
in the Einstein-de Sitter universe is
$D(t)=a(t)  \propto t^{2/3}$.\\
Let us perform  further analysys of the basic equations.
It is convenient to use the growing solution  $D(t)=a(t)$ as a new time
 variable instead
 of $t$ (e.g. \cite{GSS}, \cite{Kof})
 and introduce a comoving
velocity $\vec u ={d\vec x \over da}=\vec v /{a \dot a}$ in
 respect with this time
variable.
  Then eq.~(\ref{eq:euler}) has the form
\begin{equation}
{\partial  \vec u \over \partial a}+(\vec u \nabla_x)\vec u
 = (a\dot a)^{-2}\biggl( (3a^2 \ddot a) \vec u - \nabla_x\phi \biggr)
                \label{eq:eul1}
\end{equation}
 Let $\Phi$ be
 velocity potential so $\vec u = \nabla_x\Phi$.
  We also observe that the combination
$A=({3 \over 2}H^2a^3)^{-1}= -(3 \ddot a a^2)^{-1}$ does
 not depend on time in the
matter-dominated
 Einstein-de Sitter Universe.
 Then from eq.~(\ref{eq:eul1})
 we find
a general relation between the velocity potential and the Newtonian
gravitational potential \cite{KS}
\begin{equation}
{\partial \Phi \over \partial a}+{1\over2}(\nabla_x\Phi)^2=
= -{3 \over {2a}}  \biggl(  \Phi +A \phi \biggr).
                \label{eq:bernul}
\end{equation}
This equation is often reffered to  as the Bernoully equation.
Substituting density $\rho$ from the right hand side of eq.~(\ref{eq:poiss})
 into eq.~(\ref{eq:cont}), we  find the second equation linking
scalar potentials $\phi$ and $\Phi$
\begin{equation}
\nabla_{x_i} \biggl[{\partial  \over \partial a}(a \nabla_{x^i} \phi)+
(A^{-1}+a\nabla_x^2\phi)\nabla_{x^i}\Phi \biggr] =0.
\label{eq:link}
\end{equation}
The structure of this equation is $\nabla_x \cdot \vec \Sigma=0$, where
the vector $\vec \Sigma (\vec x) $ is the expression in the square
 brackets in
 eq.~(\ref{eq:link}). Using the Helmholtz theorem, we find
\begin{equation}
\vec \Sigma (\vec x) = \nabla_x \times \biggl[ {1 \over 4\pi}
 \int {d^3x' \over
{|\vec x - \vec x'|}}  \nabla_{x'} \times \vec \Sigma (\vec x') \biggr].
\label{eq:helm}
\end{equation}
After some vector algebra, we end up with the formula
\begin{equation}
{\partial  \over \partial a}(a \nabla_{x_i} \phi)+
A^{-1} \nabla_{x_i} \Phi = { a \over 4\pi}
 \nabla_{x^k} \nabla_{x_i}  \biggl[  \int {d^3x' \over
{|\vec x - \vec x'|}} (\nabla_{x'_k} \Phi) (\nabla_{x'}^2) \phi \biggr].
\label{eq:grad}
\end{equation}
 Thus we obtain two equations (\ref{eq:bernul}), (\ref{eq:grad})
for two fields $\Phi$ and $\phi$.
In the Newtonian theory we can construct the traceless symmetric tensors
$\sigma^N_{ij}=
\nabla_{x_i} \nabla_{x_j}\Phi - {1 \over 3}\delta_{ij}\nabla^2_x \Phi$
and
$E^N_{ij}=
\nabla_{x_i} \nabla_{x_j} \phi - {1 \over 3}\delta_{ij}\nabla^2_x \phi$.
Tensor $E^N_{ij}$ is the gravitational tidal field,
and corresponds to the Newtonian limit of the electric part of the
Weyl tensor.\\
 Taking derivatives
$\nabla_{x_i} \nabla_{x_j}$ from eq.~(\ref{eq:bernul}), we  can get
the  equation for  the Lagrangian time derivative
 along the trajectory
${ D \over D a}   \sigma^N_{ij}$ plus other local terms only.
Taking derivative $\nabla_{x_j}$ of eq.~(\ref{eq:grad}), we obtain
equation containing
the term ${ D \over D a}   E^N_{ij}$ plus other local terms
 in the left hand side, and
non-local integral term
  $\nabla_{x_j} \nabla_{x^k} \nabla_{x_i}  \biggl[  \int {d^3x' \over
{|\vec x - \vec x'|}} (\nabla_{x'_k} \Phi) (\nabla_{x'}^2 \phi) \biggr]$
in the right hand side. Apparently, this non-local term
cannot be reduced to the combination of the local terms.
It prevents us from  obtaining a closed set of Lagrangian equations
in the Newtonian theory.
However, in general relativity there is a closed set of Lagrangian equations
for the Weyl tensor and $\sigma_{ij}$ \cite{Ell}. In general relativity
$00$ component of the Einstein Eqs. (which contains
the Poisson equation as the Newtonian limit) plays the role of constraint
 equation, which is authomatically resolved for arbitrary time
$t > t_0$  if  it is resolved
 once at the initial
hypersurface $t=t_0$ and   evolution  equations are satisfied.
Recently  this  fact drew much attention
in connection with possible application to the LSS dynamics
beyond the linear theory  \cite{Matt}, \cite{Sal}, \cite{Ed}.
However its implication for the Newtonian theory
  is unclear, because  in the Newtonian limit the  Poisson
equation has to be resolved for each moment $t$.\\
In this contribution I will consider the general method of
following the time evolution of PDFs for the dynamical systems which
obey the closed set of Lagrangian equations.
In the case of Newtonian gravity, we will use an approximation where
basic equations are truncated and might be reduced to the
closed set of  equations containing full time derivatives only.\\

\section{Truncated Zel'dovich Approximation}

The dynamics of the system which obeys the multiple stream generalization
of the basic equations is  complicated and requires the N-body
  simulations.
For interesting cosmological models such as the CDM scenario, the structure
formation looks like complicated hierarchical pancaking and clustering from
very small to large cosmological scales.
However the gravitational clustering at  sufficiently large  scales
  $R$  can be
considered  in the quasilinear theory  in a single stream regime
ignoring small scale details.
   For this goal let us apply
the Zel'dovich approximation  for the smoothed  initial gravitational
potential  filtered  by the window
function $W(R)$ with filtering scale $R$:
\begin{equation}
\phi(\vec x; R)={1 \over 2\pi^2}\int_0^{\infty}d^3 \phi(\vec x')
W(|\vec x - \vec x'|; R).
\label{eq:filt}
\end{equation}
We will call this approach   the truncated Zel'dovich approximation,
 which was used
for different purposes  in
papers \cite{BC}, \cite{Kof}, \cite{ND},  \cite{KSPM}, \cite{CMS},
\cite{Kof2}.\\
 The mean space separation between galaxies is about
 $R_0 \sim 5 h^{-1} Mpc$. To generate
 the continuous field from galaxies , one has to smooth the survey with
the filter exceeding the  minimal scale $R_0$.
 At these scales the density contrast $
\delta \rho / \rho$ less than unity and quasilinear
truncated
 Zel'dovich approximation can be applied.
The approximated  Zel'dovich solution \cite{Zel} of the basic equations
 describes
the gradient mapping of the Lagrangian particles space $\vec q$ into physical
 Eulerian
space $\vec x$. The velocity field is defined by the gradient of
  the initial velocity potential   $\Phi(\vec q)$.
In this approximation the follow of particles density field evolves as
$\rho(\vec x) =\bar \rho(\vec q)~{|{\partial \vec x
\over \partial \vec q}|}^{-1}$.
 From mathematical point of view the truncated Zel'dovich approximation
means just the Zel'dovich approximation which is applied
to the truncated initial potential (\ref{eq:filt}) to ensure  being
 in the single stream quasilinear regime.
In this Section the  approximation   is formulated
in terms of an equation for the gravitational potential and its derivatives
corresponding to the tidal force.\\
In the Zel'dovich approximation $\phi=-A^{-1} \Phi$,
and  then the dynamical equation for the velocity potential is reduced to the
``shortened'' eq.~(\ref{eq:eul1}) without the right hand side:
\begin{equation}
{\partial \Phi \over \partial a}+{1\over2}(\nabla_x\Phi)^2=0. \label{eq:HJ}
\end{equation}
We  introduce the
 tensor of the velocity derivatives $S_{ij}(\vec x, t)=\nabla_{x_i} v_j$
 in the Eulerian space;
 for the potential motion it is reduced to
$S_{ij}=\nabla_{x_i}\nabla_{x_j}\Phi$.
Let $\lambda_i$ be its  eigenvalues.
The field of the  $S_{ij}(t)$-tensor evolves in time, its initial
value (in the Lagrangian space)
 coincides with the Lagrangian  deformation tensor
$D_{ij}=\nabla_{q_i}\nabla_{q_j}\Phi_0$. We  will call the $S_{ij}$-tensor
 as the Eulerian deformation tensor.
It is convenient to use
three invariants of the  $S_{ij}$-tensor:
\begin{equation}
  J_1=\lam_1+\lam_2+\lam_3;     ~~
           J_2=\lam_1\lam_2+\lam_1\lam_3+\lam_2\lam_3; ~~
           J_3=\lam_1\lam_2\lam_3 .                 \label{eq:inv}
\end{equation}
Taking the derivatives of $\Phi$ in respect with $\nabla_x$, we get
from eq.~(\ref{eq:HJ})
the following set of equations. First equation is
\begin{equation}
 {D \vec u \over Da} \equiv {\partial \vec u \over \partial a}+
(\vec u \nabla_x)\vec u =  0,
           \label{eq:eul2}
\end{equation}
which is the truncated Euler equation (cf. eq.~\ref{eq:eul1}))
describing the free streaming of particles in a comoving coordinates
( ${D \over Da}$ is
 the lagrangian derivative along the trajectory).
Further derivatives of this equation give us the dynamical
equations for the invariants of the deformation tensor field
\begin{equation}
  {DJ_1 \over Da}+J_1^2-2J_2 = 0, \label{eq:dyn1}
\end{equation}
\begin{equation}
           {DJ_2 \over Da}+J_1J_2-3J_3  =  0,   \label{eq:dyn2}
\end{equation}
\begin{equation}
                {DJ_3 \over Da}+J_1J_2  =  0.    \label{eq:dyn3}
\end{equation}
Also, we can rewrite the continuity equation in new variables.
Let $\rhos =a^3 \rho$ be the comoving density, then from eq.~(\ref{eq:cont})
we have
\begin{equation}
 {D \rhos \over Da}+\rhos J_1=0.  \label{eq:con1}
\end{equation}
The advantage of the set of equations (\ref{eq:eul2})-(\ref{eq:con1})
 is that it is the closed
system of equations which describes the dynamics of the cosmological
potential
density,  velocity and deformation tensor fields
$\Phi, \rhos, \vec u, J_1, J_2, J_3$  as function of
time $a(t)$ and the Eulerian position $\vec x$ in the Zel'dovich approximation.
In the next Section, using the system (\ref{eq:eul2})--(\ref{eq:con1}),
 we will derive and solve  the kinetic
 equation for the joint distribution
function of these fields. \\
 Let us  introduce  the principal radii of curvature
$R_i=\lam_i^{-1}$ of the
 three dimensional hypersurfaces of the potential $\Phi$.
 Then $J_i$ can be expressed through
$R_i$-th.
The solutions of eq.~(\ref{eq:dyn1})--(\ref{eq:dyn3})
 in terms of $R_i$ are  extremely simple:
\begin{equation}
R_i=R_{0i}+a,        \label{eq:Huy}
\end{equation}
the principal radii of curvature of the hupersurface of the potential
in the Zeldovich approximation linearly increase with ``time'' $ a(t)$.
It is   nothing but the  Huygens principle of geometrical  optics
(note the optical-mechanical
 similarity between
the ray propagations and particle trajectories in the Zel'dovich
 approximation  \cite{SZ}).
Using  (\ref{eq:inv}), (\ref{eq:Huy}) we obtain   the time evolution of
the fields  from their initial values ($J_{01}, J_{02}, J_{03}$)
  along with the characteristic:
\begin{equation}
J_{1}  =  {{ J_{01} -2 a J_{02} +3 a^2 J_{03}}
 \over { 1-a J_{01} + a^2 J_{02} - a^3 J_{03}   }},  \label{eq:eiga}
\end{equation}
\begin{equation}
J_{2}  =  {{ J_{02} -3 a J_{03}}
 \over { 1-a J_{01} + a^2 J_{02} - a^3 J_{03}   }},  \label{eq:eigb}
\end{equation}
\begin{equation}
J_{3}  =   {{ J_{03}}
 \over { 1-a J_{01} + a^2 J_{02} - a^3 J_{03}   }},  \label{eq:eigc}
\end{equation}
\begin{equation}
\rhos   =  {{ \rhos_0}
 \over { 1-a J_{01} + a^2 J_{02} - a^3 J_{03}   }},    \label{eq:eigd}
\end{equation}
and $\Phi(\vec q, a)=\Phi_0(\vec q)$, $\vec u(\vec q,  a)=\vec u_0 (\vec q)$.
 From the set of eqs.~(\ref{eq:eiga})--(\ref{eq:eigd})
 we can obtain the reverse formulas
\begin{equation}
J_{01}  =  {{ J_{1} +2 a J_{2} +3 a^2 J_{3}}
 \over { 1+ a J_{1} + a^2 J_{2}  + a^3 J_{3}   }},  \label{eq:eig1a}
\end{equation}
\begin{equation}
J_{02}  =  {{ J_{2} + 3 a J_{3}}
 \over { 1+ a J_{1} + a^2 J_{2} + a^3 J_{3}   }},  \label{eq:eig1b}
\end{equation}
\begin{equation}
J_{03}  =  {{ J_{3}}
 \over { 1+ a J_{1} + a^2 J_{2} + a^3 J_{3}   }},   \label{eq:eig1c}
\end{equation}
\begin{equation}
\rhos_0  =  {{ \rhos}
 \over { 1+ a J_{1} + a^2 J_{2} + a^3 J_{3}   }}.   \label{eq:eig1d}
\end{equation}
Additionally, we can directly get  the solution of
 eq.~(\ref{eq:HJ})
\begin{equation}
\Phi(a,\vec x)=\Phi_0(\vec q)+{(\vec x -\vec q)^2 \over 2a}.   \label{eq:hyp}
\end{equation}
The solution  (\ref{eq:hyp}) describes the deformations of 3D-hypersurface
of the potential $\Phi$, and is valid untill
the formation of folds of $\Phi$-hypersurface, which corresponds to caustics
(pancakes).
Taking the derivatives of $\Phi$ in respect with  $\nabla_x$ and  $\nabla_q$,
from (\ref{eq:hyp}) we get  the usual form of the Zel'dovich approximation
 \cite{Zel}: $\vec x = \vec q +a\nabla_q\Phi_0(\vec q)$.\\

\section{ Kinetic equation for the PDFs}

In the Zel'dovich approximation, we have derived the closed system of the
dynamical equations  (\ref{eq:eul2})--(\ref{eq:con1}), and found its
solutions
 which describes the evolution of the set of
the  fields  $\Phi, \rhos, \vec u, J_1, J_2, J_3$
( note that, for instance,
the set of fields  $\Phi, \rhos, \vec u $  does not obey the
 closed system of equations).
For the closed system of  dynamical equations one can directly
 derive a  so-called  `` kinetic'' equation for the joint PDF.
In the  context of the two dimensional geometric-optics problem
kinetic equation was derived in
\cite{MS}, \cite{Sa}.\\
Here we derive  the one-point joint PDF
 $W(\Phi, \rhos, \vec u, J_1, J_2, J_3)$
 of  the fields  $\Phi, \rhos, \vec u, J_1, J_2, J_3$
 in the Zel'dovich approximation.
Let us introduce an arbitrary function of these fields
 $f(\Phi, \rhos, \vec u, J_1, J_2, J_3)$. Taking the full time derivative
 ${Df \over Da}$ and using the set of equations
 (\ref{eq:eul2})--(\ref{eq:con1}),
 we get
\begin{equation}
{\partial f \over \partial a}+( \nabla_x)(\vec u f) -J_1 f  +
\rhos J_1 {\partial f \over \partial \rhos}  +
( J_1^2 -2 J_2) {\partial f \over \partial J_1} +
( J_1 J_2 -3 J_3) {\partial f \over \partial J_2}+
 J_1 J_3 {\partial f \over \partial J_3 }= 0.     \label{eq:kin1}
\end{equation}
Convolving this expression with
$W(\Phi, \rhos, \vec u, J_1, J_2, J_3)$, and using the fact that
 the function $f$ is an arbitrary one, we obtain
the sought kinetic equation for the joint PDF:
\begin{equation}
{D W \over D a} -J_1 W  -
 J_1 {\partial  \over \partial \rhos} (\rhos W)  -
 {\partial  \over \partial J_1}( J_1^2 -2 J_2)W -
 {\partial  \over \partial J_2}( J_1 J_2 -3 J_3)W-
 {\partial  \over \partial J_3 } (J_1 J_3 W) = 0.       \label{eq:kin}
\end{equation}
 Let $W_0(\Phi_0, \rhos_0, \vec u_0, J_{01}, J_{02}, J_{03})$
be an initial joint PDF of the initial fields
$(\Phi_0, \rhos_0, \vec u_0, J_{01}, J_{02}, J_{03})$ defined in the
Lagrangian space.\\
Thus, eq.~(\ref{eq:kin}) together with the given initial condition,
 describes the
time evolution of the statistics of the field in the problem.
Eq.~(\ref{eq:kin}) admits a simple analytical solution which can be obtained
applying the method of characteristics.
Using the time evolution   (\ref{eq:eiga})--(\ref{eq:eigd})
 of the fields along of characteristics,
 we
finally obtain the solution of the kinetic equation
\begin{equation}
W(\Phi, \rhos, \vec u, J_1, J_2, J_3; a)=
 \bigl( 1+ a J_{1} + a^2 J_{2} + a^3 J_{3} \bigr)^{-6}
W_0(\Phi_0, \rhos_0, \vec u_0, J_{01}, J_{02}, J_{03}), \label{eq:skin}
\end{equation}
where  we have to substitute
the expressions    (\ref{eq:eig1a})--(\ref{eq:eig1d})
 as  the arguments of the function $W_0$.\\
Formula (\ref{eq:skin}) is the main result of the paper. It describes
 the time evolution
of the joint PDF of the cosmological potential, density,
 velocity and its derivatives
fields
in the Zel'dovich approximation, from the given
(Gaussian or non-Gaussian) initial distribution
function $W_0$. \\

\section{ Evolution of the PDFs from Gaussian initial fluctuations}
\subsection{ General formalism for Gaussian initial statistics}
In this Section we study  the statistics of the continuous cosmological
 fields evolving from the initial Gaussian fluctuations, which is
apparently  the most attractive model of primordial perturbation.
On the other hand, in this case we can make   more advanced predictions
for the statistics of the matter distribution and motion.
For the sake of simplicity, we consider the joint PDF of the cosmological
 density,
velocity and deformation tensor. For the Gaussian initial conditions we write
the initial joint PDF as
\begin{equation}
 W_0( \rhos_0, \vec u_0, J_{01}, J_{02}, J_{03})=
Q_0(\vec u_0) \cdot \delta (\rhos_0-\bar \rhos) \cdot
 G_0(J_{01},  J_{02}, J_{03}). \label{eq:gauss}
\end{equation}
The  first factor is the Gaussian velocity distribution function.
The velocity dependence is factorized
because  there is no correlation  between
 $\nabla_q \Phi_0$ and
other fields involved in (\ref{eq:gauss})
 for the Gaussian fluctuations.
 The second factor is the initial
density distribution function, which corresponds to the perfectly
homogeneous density distribution $\rhos_0=\bar \rhos$.  This is just
the formal limit of the Gaussian density distribution with
$\sigma \to 0$. The third factor is the joint distribution function
of the invariant of the initial deformation tensor.
Substituting the form   (\ref{eq:gauss}) into (\ref{eq:skin}), we get the time
 evolution of the joint PDF
from Gaussian initial fluctuations.\\

\subsection{  Evolution of the velocity PDF from Gaussian initial fluctuations}

The initial PDF of the velocity $\vec u(x, b)$ for the Gaussian field is
\begin{equation}
Q_0(\vec u)d^3u= {1\over [2\pi\sigma_u^2]^{1/2} }
{\rm exp}\left[-{{\vec u_0}^2 \over 2\sigma_u^2}\right] \ ,
\quad
\sigma_u^2 =  \langle \vert \nabla_q \Phi_0 \vert ^2 \rangle \ ,
\label{eq;vstat}
\end{equation}
where $\sigma_u=\sigma_u(R)$ is the initial dispersion of the velocity
 $\vec u$, which
depends on the filtering scale $R$.\\
 From the joint PDF (\ref{eq:skin}) for the Gaussian initial condition
(\ref{eq:gauss}) we
see that the velocity dependence remains to be factorized for an arbitrary
moment of time $a(t)$. Integrating the joint PDF over  all the  arguments,
except velocity, we get the Eulerian velocity PDF:
\begin{equation}
Q(\vec u, b) =Q_0(\vec u_0)=  {1\over [2\pi\sigma_u^2]^{1/2} }
{\rm exp}\left[-{{\vec u}^2 \over 2\sigma_u^2}\right] \ .
\label{eq:vprob}
\end{equation}
Thus we find  that
the Eularian velocity PDF  $Q(\vec u)$
is {\it time-invariant} under the Zel'dovich approximation, in accordance with
paper \cite{Kof2},
where  this result was obtained by  a different method,
and  supported by N-body simulations beyond the Zel'dovich approximation.
The time-invariance  means   that the Eulerian velocity PDF
 remains Gaussian
with the dispersion defined by the linear theory.
 The distribution of the velocity field  $\vec u$  is
 isotropic.
Using eq.~(\ref{eq:vprob}), we can obtain the PDF of the physical velocity
$\vec v =a(t) {d \vec x \over d t}= a\dot a \vec u$:
it is Gaussian with the dispersion $\sigma_v(R)$ defined by the linear theory
and the smoothing filter.

\subsection{ Evolution of the density PDF from Gaussian initial fluctuations}

To calculate the Eulerian  density PDF $P(\rhos, a)$, we have to
integrate the joint PDF given by eq.~(\ref{eq:gauss}) over  all of
 its arguments
 except density
\begin{equation}
P(\rhos, a)
 = \int d^3 u ~d J_1 ~d J_2 ~d J_3~  W( \rhos, \vec u, J_1, J_2, J_3; a) \ .
\label{eq:dstat}
\end{equation}
To take the integral (\ref{eq:dstat}), we need  the joint distribution
function of the invariants of the Lagrangian  deformation tensor
$ G_0(J_{01},  J_{02}, J_{03})$. The joint distribution function
of its Lagrangian (initial) eigenvalues $\lam_{0i}$
 is given by \cite{D}
\begin{equation}
M(\lam_{01},\lam_{02},\lam_{03})=
{5^{5/2}\, 27 \over 8 \pi \sigma_{in}^6}\
(\lam_{01}-\lam_{02})(\lam_{01}-\lam_{03})(\lam_{02}-\lam_{03})\
{\rm exp}\left[-{1\over\sigma_{in}^2}
\biggl( 3J_{01}^2-{15 \over 2} J_{02} \biggr)\right] \
, \label{eq:dor}
\end{equation}
where $J_{01}, J_{02}$ are expressed through $\lam_{0i}$, and
 $\sigma_{in}= \sigma_{in}(R)$ is the initial variance of $ \rhos$,
which depends on the filtering scale $R$.
Then the joint distribution function of the invariants $J_i$   is
\begin{equation}
G(J_{01}, J_{02}, J_{03})   = {5^{5/2}\, 27 \over 8 \pi \sigma_{in}^6}\
{\rm exp}\left[-{1\over\sigma_{in}^2}
\biggl( 3J_{01}^2-{15 \over 2} J_{02} \biggr)\right].  \label{eq:sinv}
\end{equation}
Additionally, we have to integrate  (\ref{eq:dstat}) over the allowed
region in the $(J_{01}, J_{02}, J_{03}) $--space, for which  all three
eigenvalues $\lam_{0i}$ are real. To find that region is a non-trivial
task, the basic idea is outlined in \cite{Kof2}.
After that, taking
 integral over the velocity  in eq.~(\ref{eq:dstat}), we get
\begin{equation}
P(\rhos, a) = \int {d J_1 d J_2 d J_3 \over
\bigl( 1+ a J_{1} + a^2 J_{2} + a^3 J_{3} \bigr)^{6}}
 \delta (\rhos_0-\bar \rhos)
 G_0(J_{01},  J_{02}, J_{03})  \ . \label{eq:dstat1}
\end{equation}
We  have to substitute  $G_0$ from (\ref{eq:sinv})   and
 arguments $\rhos_0, J_{01},  J_{02}, J_{03}$ from
 (\ref{eq:eig1a})--(\ref{eq:eig1d})
 in this formula.
After some tedious algebra we can reduce the integral (\ref{eq:dstat1}) to the
simpler one-dimensional  integral which has to be performed
numerically
\begin{equation}
P(\rhos,a) =
 {{9 \cdot 5^{3/2} {\bar \rhos}^3} \over {4\pi N_s \rhos^3 \sigma^4}}
\int_{3 ({\bar \rhos \over \rhos })^{1/3}} ^{\infty}  d s\
e^{-{(s-3)^2 / 2 \sigma^2}}
\left( 1+ e^{-{6s/ \sigma^2}} \right)
\ \left( e^{-{\beta_1^2 / 2\sigma^2}}
             +e^{-{\beta_2^2 / 2\sigma^2}}
             -e^{-{\beta_3^2 / 2\sigma^2}}  \right)
\ , \label{eq:dprob}
\end{equation}
\begin{equation}
\beta_n (s) \equiv s \cdot \,5^{1/2} \left( {1\over2}
+\cos\left[{2\over3}(n-1)\pi
+{1\over3} \arccos \left({{54{ \bar \rhos}^3} \over \rhos s^3}
-1 \right)\right]\right) ,
\end{equation}
where the only parameter $\sigma(t)=a(t)\sigma_{in}$ is
 the standard deviation of the density
fluctuations
$\rhos/\bar\rhos$ in the  linear theory, and $N_s$ is the mean number
 of streams, $N_s=1$ in the single stream regime.
The expression (\ref{eq:dprob}) was derived by the different
method  earlier in \cite{Kof}, \cite{Kof2}.\\
 For  $\sigma \ll 1$ the expression (\ref{eq:dprob}) is
 reduced to the
Gaussian distribution
$ P(\rhos)= (2\pi\sigma^2)^{-1/2}
{\rm exp}
 {\biggl(  \bigl( {\rhos  \over {\bar \rhos}} -1  \bigr)^2
 /2\sigma^2 \biggr)} \ . $
For $\sigma \to 0$ we formally get $P(\rhos) \to \delta(\rhos -\bar \rhos)$,
in accordance with (\ref{eq:gauss}).
The   density PDF $P(\rhos, a)$ calculated from (\ref{eq:dprob})
 is plotted in \cite{Kof},  \cite{Kof2}
for different values of parameter $\sigma$.
The matter is evacuated from the underdense regions
with
 $\rhos < \bar \rhos$, resulting in voids
 which expand in time and
tend to occupy a larger fraction of volume, as well as
formation of  anisotropic collapsed dense pancakes which tend to occupy
a smaller fraction of volume. The positive density contrast can reach any
large value while the negative density contrast is restricted by $\rho \ge 0$.
Hence the probability function $P(\rhos, a(t))$, meaning the fraction of
volume with a given value of density, is expected to be very non-Gaussian
even in the quasilinear stage.
At large densities $\rhos \gg \bar \rhos$  PDF from (\ref{eq:dprob})
has the pancakes induced asymptota $\propto \rhos^{-3}$, which
affects such  values as skewness,  $S_3^{(Z)}=4$ in this approximation
instead of the actual value  $S_3 \approx {34 \over 7} -(n+3)$.
 However, the final smoothing (or the adhesion)
 $regularizes$
high-density asymptota of $P(\rhos)$, see \cite{Kof2}, \cite{BK}.
Formula (\ref{eq:dprob}) very well describes actual $P(\rhos)$
in the practically interesting region $0 < \rhos <$
several $ \bar \rhos$  ~ for $\sigma \le 1/2$.\\

\subsection{ Evolution of the deformation tensor PDF
 from Gaussian initial fluctuations}

The density and velocity PDFs were considered earlier in the literature.
Let us consider  the new statistics: PDF
 of {\it the Eulerian deformation tensor.}
The time evolution of the deformation tensor  gives us
alternative approach to the  LSS \cite{BondK}.
 To calculate its PDF, we have to integrate
the joint PDF over  all the arguments except $J_i$-th:
\begin{equation}
G(J_1, J_2, J_3; a)
 = \int d^3 u~ d\rhos~  W( \rhos, \vec u, J_1, J_2, J_3; a) \ .
\label{eq:gstat}
\end{equation}
Carrying out this integral and switching from invariant $J_i$-th back
to $\lam_i$-th, we obtain the joint distribution of the Eilerian eigenvalues
of the deformation tensor ${\partial u_i \over \partial x_j}$:
\begin{equation}
M(\lam_{1},\lam_{2},\lam_{3};a)=
{5^{5/2}\, 27 \over 8 \pi \sigma^6}\
\left[ {{ (\lam_{1}-\lam_{2})(\lam_{1}-\lam_{3})(\lam_{2}-\lam_{3})}
\over {\bigl( (1+a\lam_1) (1+a\lam_2) (1+a\lam_3)  \bigr)^5 }}
\right]\
{\rm exp}\left[-{1\over\sigma_{in}^2}
\biggl( 3J_{01}^2-{15 \over 2} J_{02} \biggr)\right] \
, \label{eq:edor}
\end{equation}
where $J_{01}$ and $J_{02}$ must be expressed through $\lam_i$
via eqs.~(\ref{eq:eig1a})--(\ref{eq:eig1d}).
In the limit $\sigma \to 0$ this distribution is reduced to the
initial Doroshkevich's distribution (\ref{eq:dor})  based on the
Gaussian statistics.  As $\sigma=a(t)\sigma_{in}$ is growing,
the Eulerian distribution $M(\lam_{1},\lam_{2},\lam_{3};a)$
rapidly departs from  the initial Lagrangian distribution,
 and   might be an interesting discriminative test
for the initial statistics.\\

\section{Properties of the density PDF}

\subsection{PDF from the Edgeworth perturbation series}

In this Section we consider some properties of the cosmological density PDF
using other approaches besides the Zel'dovich approximation.
In the case of the weakly non-linear dynamics ($\delta < 1$)
  when slight departure from the
initial Gaussian distribution is expected, one can use the general
decomposition series around the gaussian PDF \cite{KeSt}. Inspired
by the paper \cite{LH} on the weakly non-linearities
on statistical distributions in the theory of two-dimensional sea waves,
we suggest  similar decomposition for the cosmological
density $P(\delta)$ (in the three-dimensional case).
The  so-called  Edgeworth form of the Cram-Charlier series
for density distribution function reads as
\begin{equation}
P(\delta)={1 \over (2\pi\sigma^2)^{1/2}}
{\rm exp}
{\left( {\delta^2  \over 2\sigma^2} \right)}
{\biggl[1 + \sigma\cdot {S_3 \over 6} \cdot H_3
+\sigma^2 \cdot \biggl( {S_4 \over 24} H_4 + {S_3 \over 72} H_6 \biggr)+
 ... \biggr]} ,
\label{eq:Edg}
\end{equation}
where $H_n({\delta \over \sigma})$ is the Hermit polynoms.
  Numerical coefficients $S_n$ can be calculated in the perturbation
 series  \cite{Ber}. We found \cite{BK} that a few iterations
of the  expansion
(\ref{eq:Edg})  reproduce   the peak of $P(\delta)$
  in the interval
of $|\delta| \le 1/2$ around it
 relatively well  for  small parameter $\sigma \le 1/2$.
 It fails to reproduce $P(\delta)$ at
larger  $|\delta|$ because the series (\ref{eq:Edg})
 is an asymptotic expansion.
We understand that
 similar decomposition was independently suggested
in  \cite{JWACB}.\\

\subsection{Mystery of the log-normal distribution}

As it was noted in \cite{CJ}, \cite{Ham}, \cite{Kof2},
 the lognormal distribution
\begin{equation}
P(\rhos)={1 \over (2\pi\sigma_l^2)^{1/2}}
{\rm exp}
{\left[ { (\ln \rhos-\mu_l)^2  \over 2\sigma_l^2} \right]}
\cdot {1 \over \rhos}  \label{eq:log}
\end{equation}
is a good approximation of the actual $P(\rhos)$,
$\mu_l=\ln \mu-0.5\sigma_l^2$, $\sigma_l^2=\ln(1+\sigma^2/\mu^2)$.
 However, this
fit was checked for CDM model for moderate $\sigma \sim 1/2$ only.
The question arises, if the log-normal distribution is a universal
form of $P(\rhos)$ due to the non-linear dynamics of the cosmological
system, or it is just a convenient fit for particular cosmological
models in some intermediate regime?
 We  argue \cite{BK}  that the log-normal PDF is not a universal form but
 is close to the actual $P(\rhos)$ for some cosmological models
(including CDM) for moderate $\sigma$.\\
One can use the Edgworth seria (\ref{eq:Edg}) for $P(\rhos)$.
 From (\ref{eq:log}) we  find the skewness of the log-normal distribution
$S_3^{(log)}=3+\sigma^2$. On the other hand, the skewness of the filtered
density field is $S_3={34 \over 7} -(n_{eff}+3)$. To be the
distributed lognormally,
 density field has to satisfy the following equation \cite{BK}:
\begin{equation}
3+\sigma^2={34 \over 7} -(n_{eff} +3) .
\label{eq:blg}
\end{equation}
This is a  necessary (but not sufficient)
  condition for the $\sigma-n_{eff}$-dependence to correspond to the perfect
log-normal distribution. Clearly, it is not a general condition for
the density field in arbitrary  cosmological models.
 For instance, for CDM model
a rough approximation
(for interval $-2< n_{eff}< -1$)
 is $\sigma(n_{eff}) \approx 1.4 \cdot (-n_{eff} -0.85)$.
 This $\sigma(n_{eff})$
 dependence for CDM model is close to that of log-normal
distribution given by  eq.~(\ref{eq:blg}) for moderate
$\sigma \sim 0.4-0.6$ only, and departs from log-normal formula
for small and large $\sigma$.
To support this conclusion, the  systematic
comparison of $P(\rhos)$ against the log-normal distribution for different
$\sigma(n_{eff})$ is needed.
Thus,  some ``lognormalish'' features in
 the observed density distribution
mean that realistic cosmological model is close to the CDM one.\\

{\bf Acknowledgements}

I thank J.~Bond, F.~Bernardeau, E.~Bertshinger, A.~Dekel and
 M.~Longuet-Higgins
 for  useful discussions.\\

\vfill

\begin{thebibliography}{99}
\bibitem{Ber} Bernardeau, F. 1992. \apj\,{\bf 390}, L61.
\bibitem{BK}  Bernardeau, F. \& Kofman, L., 1993, in preparation.
\bibitem{BE} Bertschinger, E., 1992. in: {\it New Insights into the Universe},
Eds. Martinez, V. {\it  etal}.
\bibitem{Ed} Bertschinger, E., \& Jain, B. 1993, \apj\, submitted.
\bibitem{BondK} Bond, J.R., \& Kofman. L., in preparation.
\bibitem{BC} Bond, J.R., \& Couchman, H., 1988, in {\it Proc. 2-th
Canadian Conf. on GR and Rel. Astrophysics}, eds. Coley A.
\& Dyer C., World Scientific, p. 385.
\bibitem{BH} Bouchet, F. \& Hernquist, L., 1991. \apj\, {\bf 400 }, 25.
\bibitem{BSD} Bouchet, F., Shaffer, R., \& Davis, M., 1991.
 \apj\, {\bf 383 }, 19.
\bibitem{BDS} Bouchet, F., Strauss, M.,  Davis, M.,
Fisher, K., Yahil, A. \& Huchra, J., 1993.  \apj\, in press.
\bibitem{Bouch} Bouchet, F., Juszkiewicz, R., Colombi, S. \& Pellat, R,
    1992. \apj\,{\bf 394}, L5.
\bibitem{Br} Brandenberger, R., 1991, Physics Scripta, {\bf T36}, 115.
\bibitem{CF} Coles, P. \& Frenk, C. 1991. M.N.R.A.S\, {\bf 253}, 727.
\bibitem{CJ} Coles, P. \& Jones, B. 1991. M.N.R.A.S\, {\bf 248}, 1.
\bibitem{CMS} Coles, P., Melott, A., \& Shandarin, S. 1993.
 M.N.R.A.S\, {\bf 260}, 765.
\bibitem{Sal} Croudace, K., Parry, J., Salopek, D., \& Stewart, J., 1993,
\apj\, submitted.
\bibitem{D} Doroschkevich, A.G., 1970 {\it Astrofizica}, {\bf 6}, 581.
\bibitem{Ell} Ellis, G.F.R., 1971, in: {\it General Relativity and Cosmology},
ed. Sachs, R., NY: Academic Press, p. 104.
\bibitem{Fry} Fry, J., 1984. \apj\,{\bf 279}, 499.
\bibitem{Gor} Goroff, M., Grinstein, B., Rey, S-J. \& Wise, M. 1986.
  \apj\,{\bf 311}, 6.
\bibitem{GSS} Gurbatov, S.N., Saichev, A.I., and Shandarin, S.F., 1989.
                  M.N.R.A.S.\, {\bf 236}, 385; 1985
  {\it Sov. Phys. Doklady}, {\bf 30}, 921.
\bibitem{Ham} Hamilton, A. 1988, \apj\, {\bf 331}, L59
\bibitem{JBC} Juszkiewicz, R.,  Bouchet, F., \& Colombi, S.
    1993. \apj\ submitted.
\bibitem{JWACB} Juszkiewicz, R., Weinberg, D. Amsterdamski, P.,
  Chodorowski, M. \& Bouchet, F., 1993, \apj\, submitted.
\bibitem{KeSt} Kendall, M., \& Stuart, A. 1958. {\it The Advance Theory
 of Statistics}, Vol.1.
\bibitem{KS}  Kofman, L.A. \& Shandarin, S.F., 1988. Nature, {\bf 334}, 129.
\bibitem{Kof} Kofman, L.A., 1991, in {\it Primordial Nucleosynthesis and
Evolution of the Early Universe},  ed. K.Sato \&J. Audouze,
 Dordrecht: Kluver. p. 495.
\bibitem{Kof3} Kofman, L.A., 1991, Physics Scripta, {\bf T36}, 108.
\bibitem{KSPM}  Kofman, L.A., Melott, A., Pogosyan, D,Yu. \& Shandarin, S.F,
     1992. \apj\,{\bf 393}, 437.
\bibitem{Kof2} Kofman, L.A., Bertschinger, E.,  Gelb, J., Nusser, A.
\& Dekel, A., 1994. \apj\,{\bf 420}, 1.
\bibitem{LH} Longuet-Higgins, M., 1963, J.Fluid Mech. {\bf 17}, 459.
\bibitem{MY} Monin, A. \& Yaglom, A., 1975. {\it Statistical Fluid
  Mechanics}, MIT Press.
\bibitem{MS} Malakhov, A., \& Saichev, A. 1974. Sov. Phys.-JETP, {\bf 40}, 467.
\bibitem{Matt} Matarrese, S., Pantano, O., \& Saez, D., 1993, Phys. Rev.
 {\bf D47}, 1311.
\bibitem{O} Ostriker, J., 1988, in: {\it IAU Symp. No 130}, Eds. Audoze, J.
\& Szalay, A., Dordrecht: Reidel.
\bibitem{ND} Nusser, A.\& Dekel, A., 1990. \apj\, {\bf 362}, 14.
\bibitem{PS} Padmanabhan, T., \& Subramanian, K. 1993. \apj\, {\bf 410} 482.
\bibitem{Peeb}  Peebles P.J.E. 1980.
 {\it The Large-scale Structure of the Universe},
     Princeton University Press.
\bibitem{Sa} Saichev, A. 1976. Sov-Radiophysics, {\bf 19}, 418.
\bibitem{Sau} Saunders, W.{\it et al.}, 1991, Nature {\bf 349}, 32.
\bibitem{S} Saslaw, W., \& Hamilton, A. 1984, \apj\, {\bf 276 }, 13.
\bibitem{SZ}  Shandarin, S.F. and Zel'dovich, Ya.B., 1989. {\it Rev.Mod.Phys.},
      {\bf 61}, 185.
\bibitem{T}  Turok, N., 1991, Physics Scripta, {\bf T36}, 135.
\bibitem{Zel}  Zel'dovich, Ya.B., 1970. A\&A\, {\bf 5}84.
\bibitem{CW} Weinberg, D. \& Cole, S. 1992. M.N.R.A.S\, {\bf 259}, 652.
\end{thebibliography}
\end{document}